\documentclass[letter]{aa} 

\usepackage{graphicx}
\usepackage{txfonts}
\bibpunct{(}{)}{;}{a}{}{,}
\usepackage{hyperref}
\hypersetup{
    colorlinks=true,
    linkcolor=blue,
    filecolor=magenta,      
    urlcolor=blue,
    citecolor=blue,
}
\begin{document}

   \title{Limits on the non-thermal emission of the WR--WR system Apep}
   \titlerunning{Apep with \textit{Fermi}-LAT}

   \author{G. Martí-Devesa
          \and
          O. Reimer
          \and
          A. Reimer
          }

   \institute{Institut f\"{u}r Astro- und Teilchenphysik, Leopold-Franzens-Universit\"{a}t Innsbruck, A-6020 Innsbruck, Austria\\\email{guillem.marti-devesa@uibk.ac.at}
}

   \date{Received ---, ---; accepted ---, ---}

  \abstract
  {Colliding-wind binaries (CWBs) constitute an emerging class of $\gamma$-ray sources powered by strong, dense winds in massive stellar systems. The most powerful of them are those binaries hosting a Wolf-Rayet (WR) star. Following the recent discovery of Apep -- the closest known Galactic WR--WR binary -- we discuss the non-detection of its putative high-energy emission by the \textit{Fermi} Large Area Telescope (\textit{Fermi}-LAT) in this Letter. The limits reported in the GeV regime can be used to set a lower limit on the magnetic field pressure density within the shocked wind-collision region (WCR), and to exclude Apep as a bright $\gamma$-ray emitting binary. Given that this WR--WR system is the most luminous CWB identified until now at radio wavelengths, this result proves unambiguously that non-thermal synchrotron emission is not a suitable identifier for the subset of $\gamma$-ray emitters in this class of particle accelerators. Rather, Apep could be an interesting case of study for magnetic field amplification in shocked stellar winds.}


   \maketitle
%

\section{Introduction} \label{sec:intro}
        
Binary stellar systems with strong, interacting winds have been suggested as particle accelerators for a long time \citep[see, e.g.][]{Eichler93, Benaglia03, Pittard06, Reimer06}. These are the so-called colliding-wind binaries (CWBs), where the winds of the two components form a bow-shaped shock in the wind-collision region (WCR). Part of the kinetic energy of the winds is hence injected in the WCR, accelerating protons and electrons via diffusive shock acceleration \citep[DSA, for a review see e.g. ][]{Schure12}, which are subsequently cooled by multiple processes, including radiating $\gamma$ rays. Therefore, binaries containing a Wolf-Rayet (WR) or even early-type stars are potential $\gamma$-ray sources.
 Although many CWBs have been found to emit radio synchrotron radiation\footnote{See the updated particle-accelerating CWB catalogue at \url{http://www.astro.ulg.ac.be/~debecker/pacwb/home.html}.} \citep{Becker13}, only two of such systems (\object{$\eta$ Carinae} and \object{$\gamma^2$ Velorum}) have been detected in the GeV regime \citep[see e.g.][]{Abdo10, Werner13, Pshirkov16, Marti-Devesa20}.
 
Recently, \object{WR 70-16} (2XMM J160050.7-514245, most commonly known as Apep) was identified as the closest\footnote{See the updated Galactic WR catalogue by \cite{Rosslowe15} at \url{http://pacrowther.staff.shef.ac.uk/WRcat/}. Other WR--WR candidates have been found e.g. in the Magellanic Clouds \citep{Sana13}.} WR--WR binary \citep[$\sim 2.4$~kpc,][]{Callingham19, Callingham20, Marcote21}. Despite having slightly different stellar types (the carbon-rich and nitrogen-rich WC8 and WN4-6b, respectively), both stars have a similar terminal wind velocity ($v_{\infty , \textrm{WC}}= 2100 \pm 200$  km s$^{-1}$ and $v_{\infty,\textrm{WN}}= 3500 \pm 100$ km s$^{-1}$) from which a similar mass loss was derived ($\dot{M}_{\textrm{WC}} / \dot{M}_{\textrm{WN}} = 0.73 \pm 0.15$), providing a rather balanced, shocked WCR (opening angle of $\sim 150^{\circ}$). Even more, such a structure has been resolved at radio wavelengths \citep{Marcote21}. Assuming a standard WR mass loss \citep[e.g. $\dot{M}_{\textrm{WC8}}= 10^{-4.53}$ M$_{\odot}$ yr$^{-1}$,][]{Sander19}, Apep stands out as a rather energetic binary system -- and, in fact, it is the most luminous non-thermal radio synchrotron emitter among the known particle-accelerating CWBs \cite[$L_{\textrm{rad}}\sim 10^{31}$ erg s$^{-1}$ for an assumed distance of $2.4$~kpc,][]{Becker07, Callingham19}. Consequently, \cite{delPalacio22} made a first attempt to model the broad-band spectrum of the system, deriving the mass losses in the system and predicting the existence of leptonic inverse Compton (IC) of the stellar photon field and hadronic $\pi^0$ decay from proton-proton interactions of the stellar winds, both overlapping at GeV energies for low magnetic fields ($B_{\textrm{WCR}}\sim0.07$~G), which could be detectable by current observatories. 

In this Letter we provide observational constraints on the Apep system and, as promising as it may seem at first sight, we briefly discuss the adequacy of considering it a high-energy non-thermal emitter, as well as its impact in the emergence of CWBs as a $\gamma$-ray source class.

\section{\textit{Fermi}-LAT observations and results} \label{sec:fermi}

The \textit{Fermi} Large Area Telescope (\textit{Fermi}-LAT) is a leading facility exploring the GeV sky \citep{FermiLAT, Ajello21}. For this analysis, we used data taken by LAT in the 0.1--100 GeV range between August 2008 and April 2022 (239557417 -- 671445532 mission elapsed time, MET), in a region of interest (ROI) sized $20^{\circ} \times 20^{\circ}$ and centred on the nominal position of Apep \citep[RA~$ = 16$h $00$' $50.48$'', DEC = $-51^{\circ}$ 42' $44.98$'',][]{Callingham19}. Requesting a maximum zenith angle of $90^{\circ}$, the analysis was performed on P8R3 \texttt{SOURCE} data \citep{P8R3}. Fluxes were obtained performing a joint, binned ($0.05^{\circ}$) maximum-likelihood fit \citep{Mattox96} on the four point-spread function (PSF) quality reconstruction data quartiles (PSF0--3). We assumed a common background model (4FGL DR2 catalogue, gll\_psc\_v27), while the diffuse emission was assessed using `gll\_iem\_v07.fits` and `iso\_P8R3\_SOURCE\_V3\_v1.txt` for the Galactic and isotropic components, respectively \citep[with the corresponding instrument response functions for each quartile, see][]{4FGL}. Given that no source is present in the catalogue at the position of Apep, a new point-like source was added initially assuming a power-law spectrum with a standard spectral index $\Gamma=2$. To evaluate detection significance, we used the test statistic $TS = - 2\;\ln L_{\textrm{max},0} /L_{\textrm{max},1}$, where $L_{\textrm{max},0}$ is the likelihood value for the null hypothesis and $L_{\textrm{max},1}$ is the likelihood for the tested model. In our fit, we freed the normalisation of all sources within $7^{\circ}$ of the ROI centre, together with all parameters from sources detected with $TS > 5000$.

\begin{figure}
   \centering
   \vspace{0.2cm}
   \includegraphics[width=0.96\hsize]{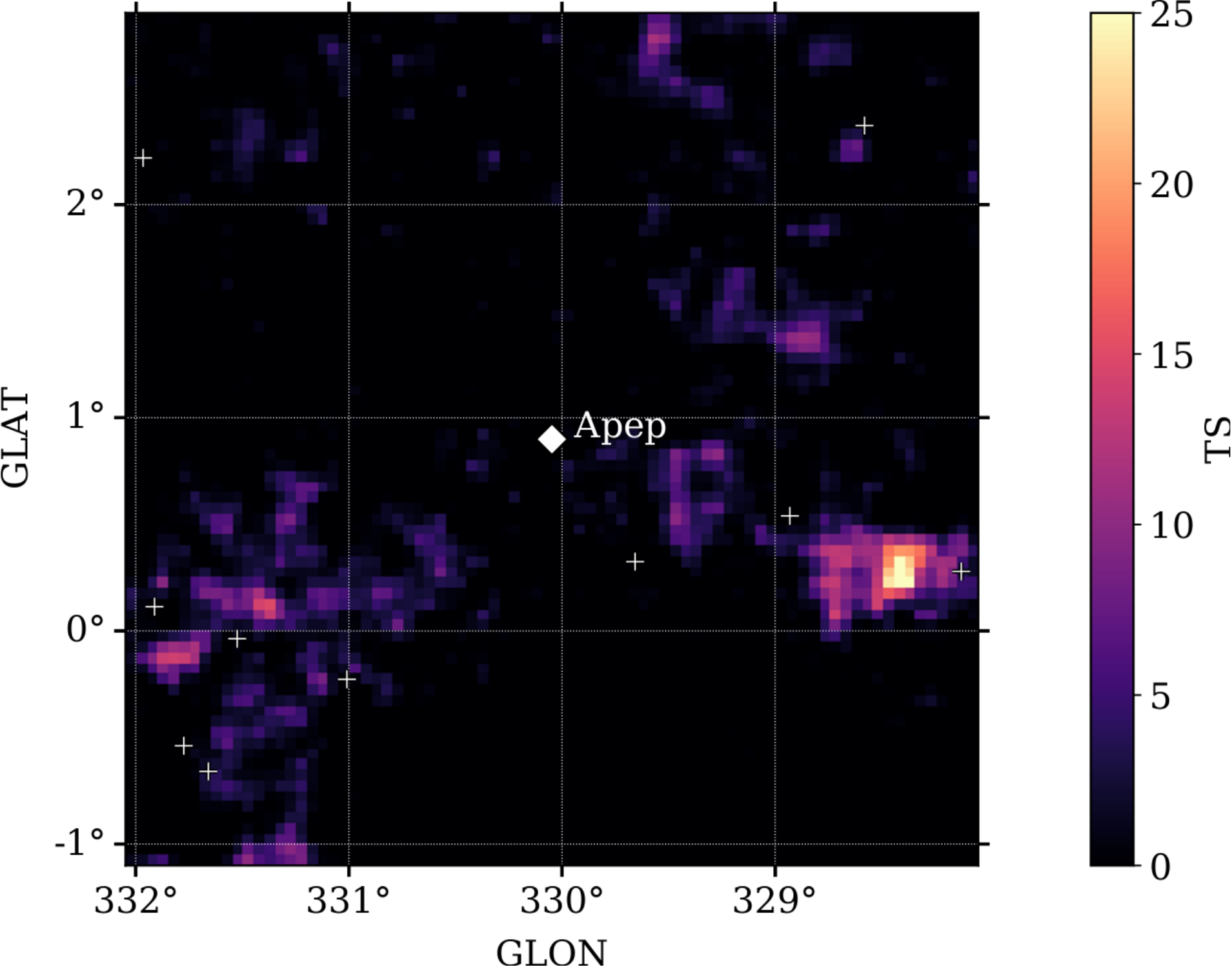}
      \caption{ Residual $TS$ map of a zoomed section of the ROI centred on Apep, also including a signal associated with the binary. White crosses represent 4FGL-DR2 sources. No significant $\gamma$-ray signal can be appreciated at the nominal position of the binary.
              }
         \label{fig:tsmap}
   \end{figure} 

We found no significant emission from the WR--WR system Apep between 0.1 and 100 GeV ($TS\sim 0$, see Fig. \ref{fig:tsmap}), setting an upper limit of $9.3 \cdot 10^{-10}$ ph cm$^{-2}$ s$^{-1}$ at a $95\%$ confidence level (equivalent to an energy flux of $1.0 \cdot 10^{-12}$ erg cm$^{-2}$ s$^{-1}$). Furthermore, we derived upper limits for the spectral energy distribution (SED, see Fig. \ref{fig:sed}). To test the effect of choosing a standard spectral index, we also assumed a more physically motivated one \citep[$\Gamma =2.3$,][]{Kowal21, delPalacio22}, providing similar results ($TS\sim 0.7$, $2.9 \cdot 10^{-9}$ ph cm$^{-2}$ s$^{-1}$, and $1.7 \cdot 10^{-12}$ erg cm$^{-2}$). If instead we use the \texttt{find\_sources}\footnote{\url{https://fermipy.readthedocs.io/en/latest/advanced/detection.html}} algorithm from \texttt{fermipy} \citep[][requiring $5\sigma$ and a minimum distance of $0.5^{\circ}$]{Fermipy}, no excess would be found within $\sim 2^{\circ}$ of the binary position, nor after a fit with the improved background model\footnote{The possible new signals detected along the ROI (14 in total) are attributed to a larger exposure in this analysis compared with the original model derivation, as well as a different methodology. A deep, conclusive study of their nature is beyond the scope of this Letter.}. No significant signal was found either employing the PS data-model deviation estimator as an alternative to explore systematic effects in our residual map derivation \citep{Bruel21}.

\section{Discussion and conclusions} \label{sec:disc}

As previously mentioned, the double WR composition of Apep makes it a perfectly pertinent candidate for $\gamma$-ray detection among the known CWBs. Hence it is a priority establishing how powerful this binary actually is. The luminosity of its winds considering the most energetic component (i.e. presumably the nitrogen-rich WR) is $P_{\rm kin} \sim \frac{1}{2} \dot{M}_{\textrm{WN}} v_{\infty , \textrm{WN}}^2 \sim 1.5 \cdot 10^{38}$ erg s$^{-1}$, which is an order of magnitude larger than any other CWB. This value relies on the extreme terminal velocity measured \citep[$v_{\infty , \textrm{WN}}=3500\pm100$ km s$^{-1}$,][]{Callingham20}. Although large wind velocities are possible for WR stars \citep[for a review see e.g.][]{Crowther07}, the derived velocity would correspond to the largest wind measured from any WN in the Milky Way or the Large Magellanic Cloud by $\sim 1000$ km s$^{-1}$ \citep{Hamann06, Hainich14}. Hence a more standard $v_{\infty , \textrm{WN}}=1500$ km s$^{-1}$ would push $P_{\rm kin}$ down to $2.8 \cdot 10^{37}$ erg s$^{-1}$, comparable to other CWBs -- including $\eta$ Carinae. This has allowed us to, in a simplistic manner, confront their efficiency $\xi_{\gamma}$ at converting wind power into high-energy radiation. Following the results from the 4FGL catalogue \citep{4FGL}, we find that the $\gamma$-ray luminosities of $\eta$ Carinae and $\gamma^2$ Velorum are $L\left( >0.1 \;{\rm GeV }\right) = 1.2 \cdot 10^{35}$ erg s$^{-1}$ ($\xi_{\gamma}\sim 4\cdot 10^{-3}$) and $L\left( >0.1 \;{\rm GeV }\right) = 6.7 \cdot 10^{31}$ erg s$^{-1}$ ($\xi_{\gamma}\sim 4\cdot 10^{-6}$), respectively. From our upper limit on the energy flux from the newly discovered binary, we were able to constrain its luminosity to $L\left( >0.1 \;{\rm GeV }\right) < 6.9\cdot 10^{32}$ erg s$^{-1}$ ($\xi_{\gamma} < 2.5 \cdot 10^{-5}$ for $v_{\infty , \textrm{WN}}=1500$ km s$^{-1}$). Consequently, we can see that despite being copiously brighter at radio wavelengths, Apep must be considerably less efficient than $\eta$ Carinae at radiating high-energy $\gamma$ rays even with conservative wind velocities. Higher $v_{\infty}$ values would establish this CWB as an inefficient system similar to $\gamma^2$~Velorum at best ($\xi_{\gamma} < 4.6 \cdot 10^{-6}$ for $v_{\infty , \textrm{WN}}=3500$~km~s$^{-1}$).

In order to get a glimpse of the possible causes, one should also note that the measured terminal velocity is noticeably larger than the expansion velocity of the surrounding dust plume of Apep ($v_{\rm plume}=570\pm70$ km s$^{-1}$), probably requiring strongly anisotropic winds to reconcile the observational results \citep{Callingham19}. This is in stark contrast with the spherically symmetric wind assumed in the derivation of the terminal velocity from the He I P Cygni profile measured with near-IR spectroscopy \citep{Eenens94, Callingham20}. As proposed by \cite{Callingham19}, a fast polar wind together with a slow equatorial one might be able to explain the anisotropy inferred. Furthermore, the radio flux evolution of the system is also preferentially modelled with an asymmetric wind expansion, which does not seem to be particularly aligned with the orbital plane \citep[][]{Bloot22}. However, regardless of the orientations of the stellar rotations axes with respect to the orbital angular momentum, the presence of anisotropies is particularly relevant for the energy gain in DSA. A particle with energy $E$ will increase it at a rate of

\begin{equation}
\dot{E}_{\rm DSA} = \frac{c_{\rm r}-1}{3c_{\rm r}\kappa_{\rm a}} V_{\rm s}^2 E,\nonumber
\end{equation}

\noindent where $c_{\rm r}$ is the compression ratio in the WCR, $\kappa_{\rm a}$ is the diffusion coefficient perpendicular to the wind contact surface, and $V_s$ is the upstream velocity of the shock, that is $V_s \approx v_{\infty , \textrm{WN}}$ in this case \citep{Schlickeiser02}. Therefore, the presence of a slow equatorial wind would significantly affect any characterisation of the intra-binary shock in Apep. As a result, it is clear that still many observational details are missing to understand this WR--WR system.

\begin{figure*}[ht!]
   \centering
   \includegraphics[width=\hsize]{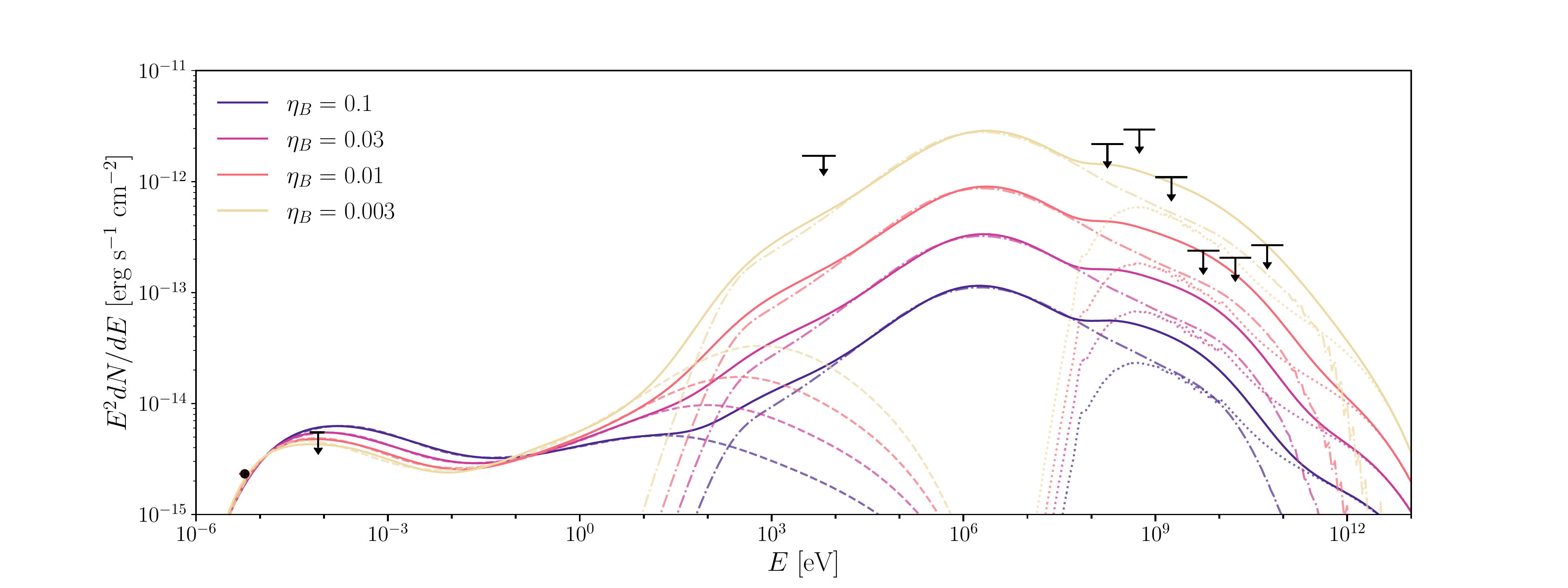}
      \caption{SED of the WR--WR binary Apep as modelled by \citet{delPalacio22} for different fractions of the magnetic field energy density. It includes synchrotron emission and thermal bremsstrahlung (dashed lines), IC radiation (dot-dashed lines), and $\pi^0$ decay (dotted lines). The total differential flux is represented by the solid lines, leaving out further IR, optical, and X-ray thermal components. In black, one can see the observational constraints set by the Australia Telescope Compact Array \citep[ATCA,][]{Callingham19}, the \textit{XMM-Newton} X-ray observatory \citep{delPalacio22}, and \textit{Fermi}-LAT (this work). The high frequency-radio (19.7 GHz) and X-ray (3--10 keV) constraints are represented as upper limits given the presence (or possible existence for radio) of thermal components. Together with the ATCA limits, our SED results are compatible with $0.01\lesssim\eta_B\lesssim0.03$ (i.e. $0.13  \; \textrm{G}\lesssim B \lesssim 0.27$~G), which is consistent with the limits derived from the integrated photon flux.}
              
         \label{fig:sed}
\end{figure*} 

Despite these limitations, a multi-wavelength modelling of Apep was performed by \cite{delPalacio22}, which did not consider anisotropic effects as a first approximation. In their approach, they employed a multi-zone model for quasi-stationary shocks \citep{Palacio16,Palacio20}, injecting a relativistic particle distribution at the WCR with a spectral index $p=2.42$ derived from radio observations. Consequently, in this parametrisation, only a fraction of the total available power was injected into the WCR, and only $2\%$ of that fraction was transferred to electrons. The most relevant result for our discussion here is their prediction of IC and $\pi^0$-decay GeV components depending on the magnetic field strength in the WCR. Our limits are not compatible with their most optimistic results (see Fig.~\ref{fig:sed}), suggesting that the magnetic field pressure as a fraction of the thermal pressure should be at least $\eta_B>0.01$ \citep[consistent with the constrains on the integrated photon flux, see also Table 2 in][]{delPalacio22}. Together with upper limits on $B$ from radio observations, this narrows the available parameter space to $0.01\lesssim\eta_B\lesssim0.03$. Such a small range restricts the efficiency in converting the available kinetic power within the WCR into cosmic rays (CRs) to $ \xi_{CR} \sim 3\cdot 10 ^{-2}$ \citep[not far from estimates on $\eta$ Carinae, see][in stark contrast with the $\xi_{\gamma}$ results]{Farnier11, White20, delPalacio22}. Thus the underlying model assumptions might foster an overconfident parameter space exclusion. Further constraints may be obtained with hard X-ray observations (e.g. with \textit{NuSTAR}; del Palacio, Private Communication).

Although we note that the limited $\eta_B$ values do not reach equipartition, the derived magnetic field limit ($0.27\; \rm{G} \gtrsim \mathnormal{B} \gtrsim0.13$ G) is already at the high end of values expected within the WCR in generic CWBs \citep[see e.g.][]{Kissmann16, Kowal21}. Relevant effects such as magnetic amplification of the initial magnetic field $B_0$ could also play an important role for high Alfv\'{e}nic Mach numbers \citep[$B_{\rm tot}^2/B_0^2\propto M_{\rm Alfven}$, where $B_{\rm tot}$ is the total magnetic field, see e.g.][]{Caprioli14a}, which might exacerbate synchrotron particle losses -- precluding $\gamma$-ray detection while still boosting bright radio emission. In the particular case of CWBs, WRs naturally provide plasma outflows energetic enough to significantly amplify the magnetic field and reach $\delta B/B_0 > 1$, since $M_{\rm Alfven} = \sqrt{4\pi \rho_{\rm w} V_{\rm s}^2}/B_0$ where $\rho _{\rm w}$ is the wind density. And indeed, after exploring particle acceleration in a generic WR--B system, \cite{Grimaldo19} concluded that strong magnetic field amplification by means of resonant-streaming instabilities was particularly relevant due to high compression ratios in the WR side of the shock \citep[see also][]{Caprioli09, Pittard21} -- hence hypothesising that this physical process is a possible mechanism behind the lack of high-energy emission in radio-emitting CWBs. Thereby, magnetic field amplification might be able to explain the low IC efficiency limit that we obtain for Apep.

That would raise the question of why no hadronic emission is observed either -- which is the preferred mechanism in $\gamma^2$~Velorum, for example \citep{Reitberger17}. For another recently studied CWB \citep[\object{HD~93129A}, see][]{Palacio20, Marti-Devesa21}, 3D magneto-hydrodynamic simulations at periastron yielded plasma densities $10^3$ times lower in its WCR than in the detected $\gamma$-ray CWBs \citep{Reitberger18}. Such densities, in turn, led to the prediction of very low GeV fluxes. Here we can estimate, analogously but in a simplified manner, the upstream particle flux at the shock boundary of Apep. The relative position $x_{\rm A}$ of the shock with respect to the most powerful of the components was determined by the balance of the wind pressures \citep[$\rho_{\rm A}v_{\rm A}^2 = \rho_{\rm B}v_{\rm B}^2$, following][]{Eichler93}. Therefore,

\begin{equation}
x_{\rm A}=\frac{1}{1+\sqrt{\eta}}\; D, \nonumber
\end{equation}

\noindent where

\begin{equation}
\eta=\frac{\dot{M}_{\rm B}v_{\rm B}}{\dot{M}_{\rm A}v_{\rm  A}} \nonumber
\end{equation}

\noindent and $D$ is the distance between the two stars. We can use such a simple description to estimate the particle density flux as
$\phi=\dot{M}_{\rm A}/(4\pi m_p x_{\rm A}^2) $
at the WCR, where $m_p$ is the proton mass. Since both stellar components are similar, the result would not vary dramatically if we were to choose the WC component instead for the Apep system. Using previously known orbital solutions and wind models for the $\gamma$-ray detected CWBs \citep[][and references therein]{Pittard02, North07, Madura12}, we have obtained that $\phi_{\rm Apep} = 1.3 \cdot 10^{-3} \phi_{\gamma^2\;\rm Vel} = 5.1 \cdot 10^{-3} \phi_{\eta\;\rm Car}$. Conservatively, we used the semi-major axis as an average stellar separation $D$ to compare it with Apep, whilst contrasting particle fluxes at periastron for both systems with HD~93129A provides similar ratios -- comparable with the density results from the magneto-hydrodynamic simulations. The low particle densities available, without considering injection efficiency into the WCR, will in turn highly impact the non-thermal emissivity in those CWBs. We reiterate that our previous calculation is a rough estimate and neglects many relevant aspects \citep[for example the aforementioned anisotropy, or a realistic radial velocity parametrisation at periastron, see e.g.][]{Lamers99}; however, it might provide clues as to the initial conditions required at the WCR to detect high-energy radiation. Beyond those, for Apep, it is particularly interesting to also note that backreaction of the particles on the shock can lead to magnetic field amplification, while substantially increasing proton energy losses \citep{Ellison95, Grimaldo19}. Hence resonant-streaming instabilities can also suppress any existing hadronic $\gamma$-ray component, not only IC. 

Moreover, neither $\eta$ Carinae nor $\gamma^2$ Velorum display a detectable synchrotron component, while Apep is the most luminous CWB at GHz frequencies. In consequence, its non-detection at $\gamma$ rays provides unequivocal evidence that non-thermal radio emission does not hold as a reliable smoking gun for non-thermal high-energy emission in CWBs, despite proving electron acceleration. In the particular case of Apep, it could be argued that the high eccentricity of the system \citep[$e\sim0.7$,][]{Bloot22} might lead to better conditions within the WCR at periastron -- at first order $L_{\gamma} \propto 1/D$ for both IC and $\pi^0$-decay processes, but this scaling cannot be generalised since it is sensitive to the particle cooling length, wind velocity profile, and further effects that also depend on the orbital geometry and stellar types \citep[see e.g.][]{Hamaguchi18, Marti-Devesa20, Pittard20}. This surely deserves further, detailed investigation solely considering the large energy budget of the system. But, unfortunately, the estimated large period of the binary ($P \sim $ 116 -- 142 yr) and previous periastron passage ($T_0 \sim$ 1926 -- 1966) will make the observation of the next periastron by \textit{Fermi}-LAT unlikely \citep{Han20,Bloot22}. Until such large uncertainties are reduced and the anisotropy of the winds is fully characterised, any high-energy model of this binary will remain inconclusive. All things considered, we shall not forget that geometrical effects significantly impact the high-energy emission of CWBs, largely restraining non-thermal radiation in systems with apparent ample wind power.

\begin{acknowledgements}
The \textit{Fermi}-LAT Collaboration acknowledges support for LAT development, operation and data analysis from NASA and DOE (United States), CEA/Irfu and IN2P3/CNRS (France), ASI and INFN (Italy), MEXT, KEK, and JAXA (Japan), and the K.A.~Wallenberg Foundation, the Swedish Research Council and the National Space Board (Sweden). Science analysis support in the operations phase from INAF (Italy) and CNES (France) is also gratefully acknowledged. This work performed in part under DOE Contract DE-AC02-76SF00515. For the present investigation we made use of \texttt{astropy} \citep[v5.1,][]{Astropy}, \texttt{fermipy} \citep[v1.0.1,][]{Fermipy}, \texttt{Fermitools} \citep[v2.0.8,][]{Fermitools}, and \texttt{numpy} \citep[v1.21.5,][]{Numpy}. The authors also would like to thank S.~del~Palacio for providing the model SED and the insightful discussions on the Apep system following the 7th Heidelberg International Symposium on High-Energy Gamma-Ray Astronomy, together with G. Gonz\'{a}lez-Tor\`{a} for her comments on the spectral characterisation of stellar winds, and H. Fleischhack for her annotations on this manuscript.

\end{acknowledgements}

%
%

\bibliographystyle{aa}
\bibliography{sample}

\end{document}